# From Monomers to Self-Assembled Monolayers: The Evolution of Molecular Mobility with Structural Confinements

Alexandre Dhotel,[ab] Ziguang Chen,[b] Jianing Sun,[c] Boulos Youssef,[a] Jean-Marc Saiter,[a] Andreas Schönhals,[d] Li Tan,[*b] Laurent Delbreilh,[*a]

**Abstract**

The effect of structural constriction on molecular mobility is investigated by broadband dielectric spectroscopy (BDS) within three types of molecular arrangements: monomers, oligomers and self-assembled monolayers (SAMs). While disordered monomers exhibit a variety of cooperative and local relaxation processes, the constrained nanodomains of oligomers and highly ordered structure of monolayers exhibit much hindered local molecular fluctuations. Particularly, in SAMs, motions of the silane headgroups are totally prevented whereas the polar endgroups forming the monolayer canopy show only one cooperative relaxation process. This latter molecular fluctuation is, for the first time, observed independently from other overlapping dielectric signals. Numerous electrostatic interactions among those dipolar endgroups are responsible for the strong cooperativity and heterogeneity of the canopy relaxation process. Our data analyses also revealed that the bulkiness of dipolar endgroups can disrupt the organization of the monolayer canopy thus increasing their ability to fluctuate as temperature is increased.

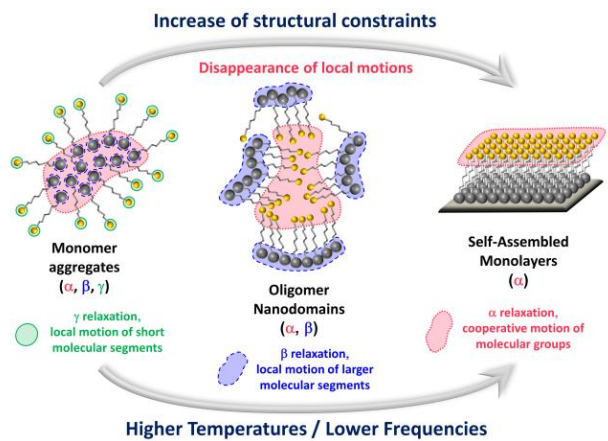

## Introduction

Molecular mobility is a crucial factor of concerns for the development of materials requiring structural flexibility[1–5] or demanding sophisticated performances at the molecular level, such as in actuators and sensors.[6–11] However, continuous shrinking of material structures and increasing number of interfaces may drastically reduce the ability of molecules to fluctuate, or even vanish thermal glass transition phenomena due to molecule immobilization.[12,13] An exemplar representation of this tendency is reflected by the incessant development of active and stimuli-responsive material systems, supramolecules, and self-assembled monolayers (SAMs).[14–17] Indeed, high molecular ordering, preparation ease, tunable functionality, and low-cost processing led these active systems to a variety of advanced applications, including electronics,[18–21] lithographic patterning,[22,23] modification of surface properties,[24–27] and photonic sensors.[9,28–30]

[a]AMME—LECAP, EA4528, International Lab., Av. de l'Université, B.P. 12, Normandie Univ. France, Université and INSA Rouen, 76801 Saint Etienne du Rouvray, France. E-mail: ltan4@unl.edu; laurent.delbreilh@univ-rouen.fr
[b]AMME—A-TEAM, Department of Mechanical and Materials Engineering, University of Nebraska, Lincoln, Nebraska, 68588, USA
[c]J. A. Woollam Co., Inc., 645 M Street, Lincoln, NE, 68508, USA
[d]BAM Federal Institute for Materials Research and Testing, Unter den Eichen 87, D-12205 Berlin, Germany
† Electronic supplementary information (ESI) available. See DOI: 10.1039/c4sm01893a

The mobility of building blocks forming such systems is often restricted due to steric hindrance and molecular interactions. Only few experimental techniques, including nuclear magnetic resonance (NMR) and scanning tunneling microscopy (STM), allow the investigation of molecular motions within such nanoscale systems. However, data acquisition rate and operating principles of these techniques limit their capabilities to map out the motions of molecules over a wide range of temperatures and within specific structural organizations. In this context, broadband dielectric spectroscopy (BDS) is an efficient complementary tool due to its high-sensitivity and broad temperature/frequency range. This allows probing multiple dipolar fluctuations in a variety of structures, such as local molecular fluctuations at low temperatures ($\gamma$ – or $\beta$– relaxation) or the dynamic glass transition ($\alpha$–relaxation) due to cooperative molecular motions at high temperatures.

In this work, spatial fluctuations of molecules within three different molecular systems are investigated: disordered monomers, oligomers forming nanophased domains, and highly ordered self-assembled monolayers (SAMs). These three structures were selected as they represent three systems with increasing structural constraints, and thus allow us to study the evolution of molecular mobility as the apparent confinement is increased.

Three different organosilane molecules, two monopolar and one bipolar, were selected as building blocks to form aforementioned three types of structures with hierarchical constrictions. Monopolar molecules, octadecyltrichlorosilane (OTS) and dodecyltrimethoxysilane (DTMS), have a silane-containing headgroup and a methyl-terminated nonpolar alkyl tail. The bipolar molecule, 11-bromoundecyltrimethoxysilane (BUDTMS), is similar to DTMS but has an extra bromine unit at the alkyl chain termination. Locations of these dipolar groups were specially selected to investigate the fluctuations of molecular segments in two specific spatial regions of SAMs: the molecule-substrate interface also called anchoring zone, and the monolayer canopy formed by molecule endgroups at the monolayer-air interface. While molecular fluctuations within SAMs have been investigated by other groups earlier,[31,32] the responses arising from these distinct regions were not explored.

## Experimental Section

### Materials

Octadecyltrichlorosilane (OTS, ≥ 95%), n-dodecyltrimethoxysilane (DTMS, ≥ 95%), and 11-bromoundecyltrimethoxysilane (BUDTMS, ≥ 95%) were respectively purchased from Acros Organics, Alfa Aesar, and Gelest, Inc. The chemical structures of monomers can be found in supporting information (Fig. SI-1). Chlorobenzene and toluene were obtained from VWR International. All chemicals were used as received without further purification. Silicon wafers, P(100) 10-20 ohm-cm and P(100) 1-10 ohm-cm, were purchased from University Wafer and Siltronix. After being cut into rectangular pieces (1.0 x 2.0 cm²), they were rinsed with Milli-Q water, and cleaned sequentially by sonication for 15 min each in ethanol and acetone. Afterwards, they were copiously rinsed with Milli-Q water and dried under a stream of nitrogen. For SAM deposition, wafer and electrode surfaces were hydroxylated by a UV/ozone treatment for 30 min, copiously rinsed with Milli-Q water and then dried under a stream of nitrogen. Interdigitated electrodes were purchased from Novocontrol Technologies GmbH (BDS1410-20-150). These electrodes were cleaned similarly as silicon wafers while contact parts were protected during the cleaning process.

### Sample preparation

Samples of alkylsilane monomers were directly deposited onto cleaned electrodes without further treatment and analyzed immediately after deposition. Oligomers of OTS, denoted as o-OTS, were prepared by depositing OTS monomers onto cleaned electrode sensors and placed in a desiccator containing a water-filled beaker at room temperature for 3 days. The high humidity in the desiccator causes the hydrolysis of the highly reactive chlorosilanes and the formation of hydrochloric acid. This leads to the subsequent condensation of monomer headgroups. Films were then kept under nitrogen atmosphere at 390 K for 2 h to complete the condensation process and remove condensation byproducts, such as water and hydrochloric acid. Rapid quenching of films to 110 K before analysis ensured a disordered structure of films. Oligomers of DTMS and BUDTMS, denoted as o-DTMS and o-BUDTMS, were prepared by adding alkylsilane precursors to a mixture of tetrahydrofuran (THF), water, and hydrochloric acid according to the following molar ratio - alkylsilane precursors : THF : H$_2$O : HCl = 1 : 50 : 20 : 0.5. Low THF, high water and HCl ratios were chosen to ensure a rapid hydrolysis and condensation of methoxysilane functionalities as well as to prevent the formation of ordered layered aggregates. Mixtures were then stirred at 500 RPM at room temperature for 24 h before dropcasting solutions onto cleaned electrode sensors. Films were then thermally treated under nitrogen atmosphere at 390 K for 2 h to complete the condensation process and remove remaining preparation compounds (THF, water, HCl) as well as condensation byproducts such as methanol. A totally disordered structure of films was ensured by rapid quenching of films to 110 K before analysis. The deposition conditions of self-assembled monolayers were developed by modifying the procedure reported by Ito et al. for the preparation of SAMs of alkyltrimethoxysilane and alkytrichlorosilane molecules.[33] After dispersing 3 mM of precursors in chlorobenzene, the solution was dispensed onto UV/ozone treated wafers or electrode sensors and placed in a desiccator containing a water-filled beaker at room temperature for 24 h. Samples were then thoroughly rinsed with ultrapure water and sonicated in toluene for 15 min. Multilayers and large molecular aggregates were removed by wiping the surface with a toluene-soaked swab. Finally, samples were sonicated in toluene for 15 min, rinsed with Milli-Q water and dried under a stream of nitrogen. A further annealing process at 330 K for 1 h under a nitrogen atmosphere has no further effect on the monolayer structure.

### Broadband dielectric spectroscopy (BDS)

Dielectric spectroscopy experiments were performed using high quality interdigitated electrodes (BDS1410-20-150) from Novocontrol Technologies GmbH (accuracy in tan(δ) ≈ 0.001, sensor diameter 20 mm, combs - gold plated copper). The comb fingers are 150 μm in width, 35 μm in thickness and spaced by 150 μm. Each electrode was calibrated prior to sample deposition by determining their respective geometric (empty) capacity ($C_0$) and substrate capacity ($C_{su}$) through measurements of a reference material (mineral B-oil from Vacuubrand) of known permittivity. For samples thicker than $d$ = 300 μm, it is assumed that the electric field penetrates only the sample and the substrate, thus the measured capacity, $C_m^*$, is given by:[34]

$$C_m^* = C_0(\varepsilon_s^* + \varepsilon_{su}^*) \quad (1)$$

where $\varepsilon_s^*$ is the complex permittivity of the sample and $\varepsilon_{su}^*$ that of the substrate. For sample thicknesses lower than $d$ = 300 μm, the investigated system can be separated into two capacitors, one of thickness $d_s$ corresponding to the sample film and a second one related to the air layer above the sample with a permittivity of 1 and a thickness of $d - d_s$. Under these assumptions the measured capacity ($C_m^*$) is:[34]

$$C_m^* = C_0 \left(\varepsilon_s^* \frac{d_s}{d} + \frac{d - d_s}{d} + \varepsilon_{su}^*\right) \quad (2)$$

Measurements were carried out in a frequency range of $10^{-1}$ to $10^6$ Hz by an Alpha-A analyzer from Novocontrol Technologies GmbH. A Quatro Cryosystem (Novocontrol Technologies GmbH) was used to control the temperature with a stability of ± 0.2 K. The temperature was increased from 110 to 440 K by successive steps of 2.5 K.

### Atomic force microscopy (AFM)

A Dimension 3100 SPM atomic force microscope was used at room temperature to obtain topographic images of SAMs. Images were recorded with a maximum resolution of 512 lines employing a tapping mode.

### Wetting tests

Static water contact angles (WCAs) were measured with a laboratory made goniometer. Drops of milli-Q water with a volume of 1 μL were deposited on monolayer surfaces and contact angles were measured using the WinGoutte software. Given values are arithmetic averages of 10 measurements performed on each sample.

### Ellipsometry

The thicknesses of monolayers were determined using spectroscopic ellipsometry (J.A. Woollam M-2000DI). Data from 193 to 1690 nm were analyzed using a three-layer model (silicon substrate/native silicon oxide/SAM). The optical constants of the silicon substrate and native oxide were fixed according to literature values.[35] To decrease the number of free fit parameters, the thickness of the native oxide layer was determined from an uncoated silicon wafer to 1.75 nm. This value was then fixed in modeling SAM layers. The refractive index of SAM layers was described using the Cauchy dispersion equation:

$$n(\lambda) = A + \frac{B}{\lambda^2} + \frac{C}{\lambda^4} \quad (3)$$

where A, B and C are constants fixed to 1.45, 0.01 μm² and 0 μm⁴, respectively, and λ is the wavelength. At least 3 measurements were performed on optically smooth and homogeneous locations.

### X-ray diffraction (XRD)

X-ray scattering patterns were obtained at room temperature on a Bruker-AXS D8 Discover diffractometer using a Cu Ka (1.544 Å) radiation. Digital data were recorded from $2\theta = 3°$ to 30° at an angular resolution of 0.04° and angular velocity of 4°.min⁻¹ on oligomer films casted on a glass substrate.

### Temperature modulated differential scanning calorimetry (TMDSC)

TMDSC experiments were conducted at a heating rate of 0.5 K.min⁻¹ with a period of 60 s and temperature amplitude of 1 °C.

Samples with masses between 10 and 18 mg were analyzed. Calorimeters were calibrated in temperature and specific heat capacity using the melting of indium. Sapphire was used to calibrate capacity signals. A Q100 calorimeter (TA Instruments) with a refrigerated cooling system (RCS) was used for measurements on oligomers. A Q200 calorimeter (TA Instruments) with a liquid nitrogen cooling system (LNCS) was used for measurements on monomers. An additional calibration in temperature with cyclohexane was required for measurements on monomers. All experiments were performed under a nitrogen atmosphere. All samples were quenched from their liquid phase to 183 K in order to promote the formation of an amorphous structure. MT-DSC curves are shown in supporting information (Fig. SI 2-3).

## Results and discussion

The high sensitivity and broad accessible frequency range makes BDS a powerful tool to study molecular motions and charge transport within a wide range of macro-, micro- and even nanoscopic materials.[36] Particularly, this technique is frequently used to investigate relaxation processes in glass-forming materials,[37,38] polymers,[39–44] and is also increasingly employed when studying structural confinement.[45–51] It can also be valuable when studying structural transitions within highly-ordered liquid and plastic crystals.[52–54]

### Data analysis

Figure 1a shows the dielectric loss ($\varepsilon$") of OTS monomers as a function of frequency and temperature in a 3D representation. Peaks observed in the dielectric loss indicate relaxation processes due to fluctuations of molecules within the sample. The conventional method to analyze such measurements is to use the model function of Haviriliak - Negami (HN-function, eq. 4) to the data in the frequency domain[55]

$$\varepsilon_{HN}^*(f, T = const.) = \varepsilon_\infty + \frac{\Delta\varepsilon}{\left(1 + \left(i\frac{f}{f_{HN}}\right)^\beta\right)^\gamma} \quad (4)$$

Where $f_{HN}$ is a characteristic relaxation frequency related to the frequency of maximal loss $f_{max} = 1/2\pi\tau_{max}$ ($f_{max}$ and $\tau_{max}$ are the relaxation rate and time, respectively) of the relaxation process under consideration, $\Delta\varepsilon$ is its dielectric strength. $\beta$ and $\gamma$ ($0 < \beta; \beta\gamma \leq 1$) are fractional parameters determining the shape of the relaxation spectra. As some of the examined samples exhibit only a weak dielectric response, the HN-function could not always be used to analyze the dielectric loss unambiguously. For this reason, experimental data have been analyzed in the temperature domain and Eq. 5 was fitted to spectra at a fixed frequency, $f$.[56–58] This model uses Gaussians to describe the dielectric loss of each relaxation process and is expressed as follows:

$$\varepsilon''(T) = \sum_{i=1}^{n} a_i \exp\left[\frac{-(T - T_i^{max})^2}{\omega_i}\right] + \left[\frac{\sigma_\infty}{\varepsilon_0 (2\pi f)^{mT+n}} \exp\left(-\frac{A}{T - T_0}\right)\right] + \chi \quad (5)$$

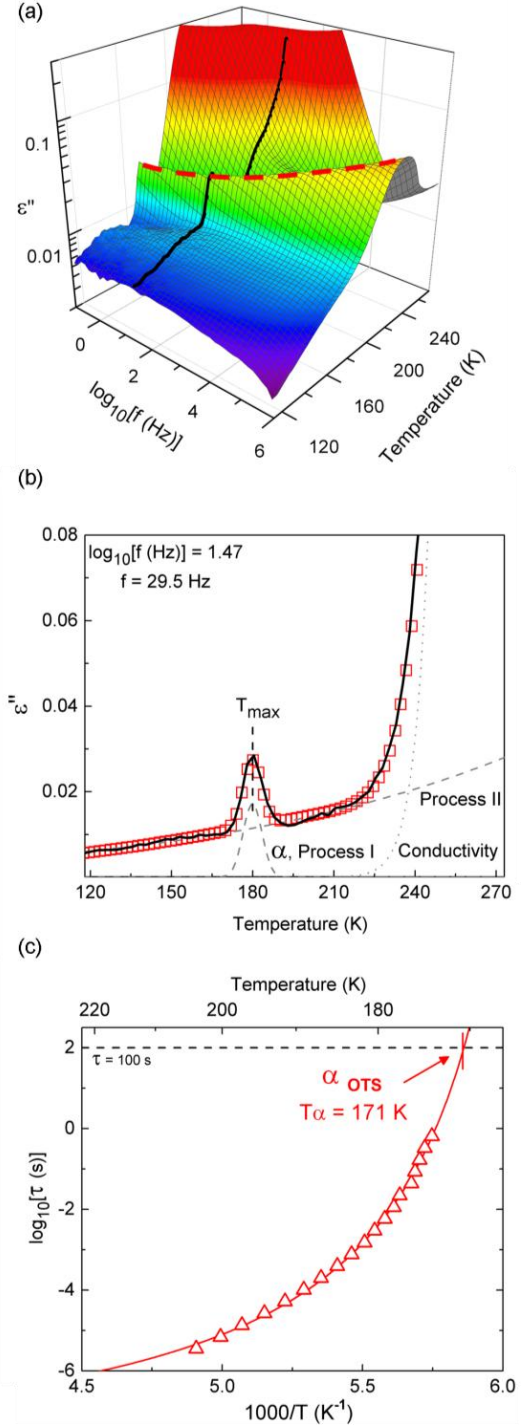

**Fig 1.** (a) Dielectric loss ($\varepsilon$") as a function of frequency and temperature for the OTS monomers. (b) Dielectric loss ($\varepsilon$") as a function of temperature at a fixed frequency of 29.5 Hz (black line in (a)). Dashed grey curves are Gaussians indicating the individual relaxation processes and the dotted grey curve represents the conductivity contribution. The red squares correspond to the resulting fit to data including all contributions (c) Relaxation time versus inverse temperature (relaxation map), the red solid curve represents the VFTH fit to experimental points.

Where $i$ counts the number of relaxation processes, $a_i$ and $T_i^{max}$ denote the amplitude and maximum position of the Gaussians; $\omega_i$ corresponds to the width of the peak when its intensity has decreased to 1/e of its maximum value. $\sigma_\infty$, $A$ and $T_0$ are parameters describing the conductivity dependence to the VFTH equation.[59–61] $m$, and $n$ are used to describe the temperature dependence of the conductivity exponent. Finally, $\chi$ is an offset. For a better accuracy of the fits, an additional broad Gaussian contribution was found to be necessary as a background contribution. The substantial width of this background contribution prevented any further analysis. Figure 1b gives an example of this analysis for the case of OTS monomers at a frequency of 29.5 Hz.

This procedure results in data pairs ($T_i^{max}$, $f$) that are used to construct relaxation map for each sample and to analyze the temperature dependence of relaxation times for each process (Fig. 1c). The simplest model to describe the temperature dependence of the relaxation time is the Arrhenius equation (Eq. 6):[32,62,63]

$$\tau = \tau_0 \exp\left(\frac{E_a}{RT}\right) \quad (6)$$

where $R$ is the general gas constant, $\tau_0$ a pre-exponential factor and $E_a$ the activation energy. This equation can be applied to localized molecular fluctuations taking place in a double wall potential. Relaxation times obeying a super-Arrhenius temperature dependence law are expected to correspond to cooperative motions characteristic for glassy dynamics. In this case, experimental data can be described by the Vogel-Fulcher-Tammann-Hesse (VFTH) equation (Eq. 7):[59–61]

$$\tau = \tau_0 \exp\left(\frac{DT_0}{T - T_0}\right) \quad (7)$$

Where $T_0$ is the Vogel temperature. $D$ and $\tau_0$ are constants. A dielectric glass transition temperature ($T_\alpha$) can be estimated by the common convention considering the temperature at which the relaxation is $\tau = 100\ s$ or $\log[\tau(s)] = 2$.[59–61]

Unless otherwise stated, all spectra obtained from dielectric experiments were analyzed as described above. Here, only relaxation maps of the investigated samples are presented in the following sections of this paper. Corresponding 3D dielectric loss vs. frequency and temperature plots can be found in Supporting Information (Fig. SI 4-6).

### Molecular dynamics of monomers

Pristine monomers were analyzed individually without further preparation. Samples were quenched to 110 K before analysis to limit the formation of ordered aggregates. However, according to the amphiphilic nature of monomers, intermolecular interactions may lead to a nanostructure where silane headgroups tend to be packed. Consequently, monomer endgroups would form the external shell of these inverted micelle-like aggregates.

Dielectric data for pristine building blocks revealed that all three monomers exhibit relaxation processes indicating different motion processes. Relaxation times of processes located at higher temperatures (lower frequencies) show a curved temperature-dependence when plotted as a function of 1/T (Fig. 2). These data can then be described by the VFTH equation that is believed to be characteristic for cooperative glassy dynamics, thus, these processes are denoted as α relaxation. Considering the similarities in the chemical compositions of monomers, it can be assumed that these processes originate from fluctuations of the dipolar alkylsilane headgroups (i.e. trichlorisilane for OTS and trimethoxysilane for BUDTMS and DTMS). The strong intermolecular interactions between these dipolar groups significantly increase their spatial correlation and cause the cooperative nature of this relaxation process. For OTS and BUDTMS monomers, the α relaxation is shifted to higher temperatures that is likely caused by the chemical composition of their alkyl chains. Probably, the long alkyl chain of OTS molecules ($C_{18}$) and polar bromine-termination of BUDTMS monomers have increased intermolecular interactions through van der Waals and electrostatic forces. All of these inherent structural constrains raise the amount of energy required for the motion of molecular segments, thus increasing the relaxation temperatures.

To correlate these α relaxations to dynamic glass transitions, TM-DSC experiments were conducted on monomers (supporting information, SI-2). All samples show a heat capacity step corresponding to their glass transitions.

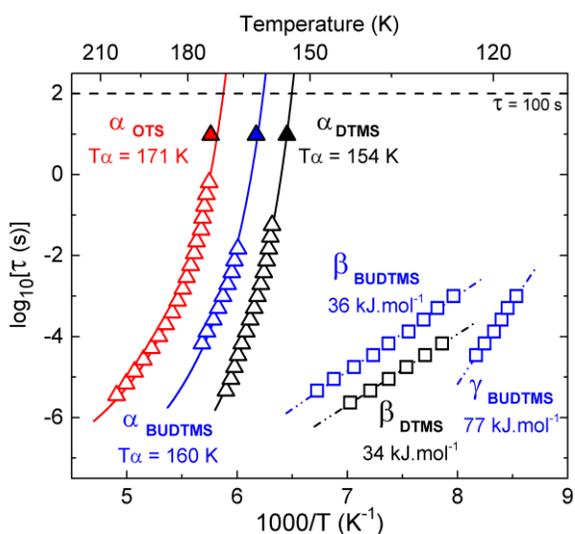

**Fig 2.** Relaxation map for the α (cooperative motions), β and γ (localized fluctuations) processes of OTS (red), BUDTMS (blue), and DTMS (black) monomers. Solid curves represent fits of the VFTH equation to the α relaxation. Dashed lines are fits of the Arrhenius formula to the β and γ relaxations. Filled points are $T_g$ values calculated from MT-DSC measurements (period = 60 s, τ = ~ 10 s).

Nevertheless, OTS monomers exhibit a more important heat capacity variation at the glass transition than DTMS and BUDTMS. The reason of this discrepancy might be the higher crystallinity ratios of DTMS and BUDTMS systems. Even though all samples were rapidly quenched before analysis to prevent molecules to crystallize, portions may have organized. The shorter alkyl chains of DTMS and BUDTMS precursors likely facilitate molecules to crystallize, thus reducing the amplitude of the glass transition phenomenon. On the contrary, the long chain of OTS limits the crystallization process and promotes the formation of amorphous phase resulting in a more pronounced glass transition. Nevertheless, glass transition temperatures estimated from TM-DSC for all monomers are in close agreement with dielectric spectroscopy measurements.

The low-temperature β relaxations correspond to localized motions of short molecular segments. As depicted in Fig. 2, only DTMS and BUDTMS monomers exhibit a β relaxation. The similarity of corresponding activation energies (ca. 35 kJ.mol$^{-1}$) indicates that they probably originate from related molecular units. As both BUDTMS and DTMS monomers carry similar trimethoxysilane (TMS) headgroups, the β relaxations observed for these compounds can be assigned to local motions of dipolar units within the TMS headgroup. This assumption is further supported by the fact that the OTS molecule, being missing such a headgroup, does not show a β relaxation.

The third relaxation process (γ) observed for BUDTMS monomers at even lower temperatures can be ascribed to the motions of bromoalkyl dipoles. The Arrhenius-like temperature dependence of relaxation times indicates localized fluctuations of these dipoles and an activation energy of 77 kJ.mol$^{-1}$.

**Molecular motions of oligomers**

o-OTS, o-BUDTMS and o-DTMS oligomers were prepared from OTS, BUDTMS and DTMS monomers, respectively. Under acidic or aqueous conditions, the trifunctional headgroups of organosilane molecules such as OTS, DTMS and BUDTMS undergo hydrolysis and subsequent condensation reactions leading to the formation of polysiloxane hyperbranched, linear, or cyclic oligomers.[64–66] Such oligomers were thus prepared from our monomers under both acidic and aqueous conditions to ensure the formation of siloxane main chains to which are attached lateral alkyl chains similar to the initial monomers. From a thermodynamic point of view the siloxane main chains and pendant alkyl chains are not mixable. Therefore, nanophase separations may occur, thus leading to the formation of alkyl chain rich nanodomains within a siloxane matrix, as already observed for poly(n-alkyl methacrylates) and poly(n-alkyl thiophenes).[67–70] This potential nanostructure is supported by X-ray diffraction patterns of oligomers (Fig. 3) which are in good agreement with the nanophase structure proposed by Beiner *et al.*.[67] In such systems, two distinct sets of diffraction peaks are observed: one is intrinsically related to the nanodomain organization and is not dependent on the length of the side chain; another, strongly dependent on the side chain length, shifts toward the high d-spacing region when the length of side chain is increased. As shown in Fig. 3, our oligomer samples exhibit a similar structure. $\theta_I$ peaks, corresponding to a d-spacing of 1.7 nm, are side chain independent and present in all solid oligomer samples. On the contrary, $\theta_\delta$ and $\theta_O$ peaks seem to be directly related to the length of alkyl chains as their d-spacing values (d = 2.3 nm for $\theta_O$ and 1.2 nm for $\theta_\delta$) are closely equivalent to the length of the corresponding monomers. At higher angles, a broader peak ($\theta_{II}$) is observed. Its corresponding d-spacing (d = 0.42 nm) is often related to the spacing between alkyl chains in highly ordered aggregates made from similar amphiphilic molecules.[64,66]

The broadness and lower intensity of the $\theta_{II}$ peak as well as the absence of other sharp diffraction peaks in o-BUDTMS indicate its isotropic organization due to its liquid state at the ambient temperature.

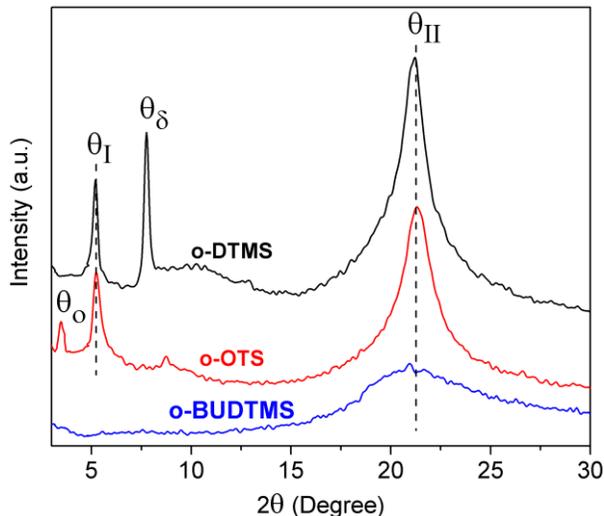

**Fig 3.** X-ray diffraction patterns of o-BUDTMS (blue), o-OTS, (red) and o-DTMS (black). O-BUDTMS being liquid at ambient temperature, only a broad scattering peak is observed indicating a disordered organization of molecules. $\theta_I$ and $\theta_{II}$ correspond to d-spacing of 1.7 and 0.42 nm and are present in both solid samples (DTMS and OTS). $\theta_O$ and $\theta_\delta$ peaks correspond to d-spacing equivalent to the length of the corresponding monomers.

It has been reported that polymers with similar nanophase separations may undergo two distinct glass transitions corresponding to the cooperative relaxations of the polymer backbone and nanodomains formed by alkyl side chains.[68] The latter relaxation, named polyethylene (PE)-like glass transition, has been found to be closely related to the average number of alkyl carbon atoms per side chain but only weakly dependent on the polymer backbone composition.

As depicted in Fig. 4, o-BUDTMS and o-OTS films exhibit both an α and a β relaxation, whereas the o-DTMS film show only a local β process at low temperatures. The observation of β relaxations in all oligomers as well as the similarity of their activation energies led to the conclusion that these β processes are related to local fluctuations of the polysiloxane chains.[65,71] While β relaxations of o-BUDTMS and o-DTMS have similar activation energies, a slightly higher value is estimated for the β relaxation of o-OTS. This difference might be due to a different polymerization process of the monomers caused by the higher hydrolysis and condensation rates of trichlorosilane headgroups in OTS molecules compared to the trimethoxysilane headgroups of BUDTMS and DTMS.[72]

An α relaxation can be clearly distinguished at higher temperatures from dielectric loss vs. temperature and frequency 3D plot of condensed OTS film (see supporting information, SI-4b). As discussed above, oligomer films are similar to macromolecular systems where the siloxane main chains form a matrix surrounding nanodomains of alkyl side chains. In such a case, the α relaxation signal of o-OTS likely originates from the fluctuations of side chains within such nanodomains. Moreover, a transition temperature of 246 K was determined by extrapolating the VFTH fit to $\tau = 100\ s$ which is between the glass transition temperatures of bulk amorphous polyethylene (PE) and that of the amorphous regions of PE constrained by crystalline regions.[73,74] Similar relaxations of alkyl nanodomains are also expected for o-DTMS and o-BUDTMS, however a weak sample response and overlapping relaxation processes prevented to assuredly distinguish such PE-like relaxations in these samples (see supporting information, SI-5 and -6). Nevertheless, o-BUDTMS exhibits a non-linear α relaxation with a transition temperature ($T_\alpha$) of 211 K. The high intensity and low temperature of this relaxation (see supporting information SI-6) led us to ascribe it to the cooperative motions of bromoalkyl groups at the termination of side chains. Indeed, by taking into account the phase-separated structure of films, it can be assumed that side chains of o-BUDTMS oligomers can form nanodomains when chains favorably interact through strong dipole-dipole interactions among bromoalkyl endgroups. Such intermolecular interactions significantly increase the cohesion between side chains within nanodomains, leading bromoalkyl dipoles to move cooperatively when the temperature of the system is increased. TM-DSC experiments were also performed on oligomers (supporting information, SI-3).

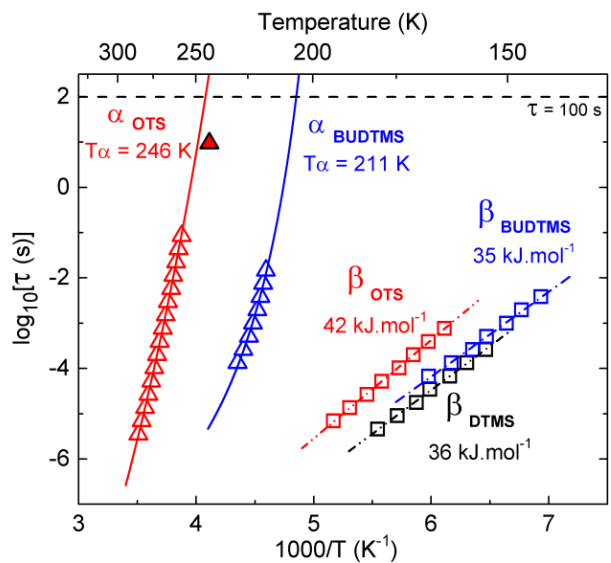

**Fig 4.** Relaxation map for the α and β relaxations of o-OTS (red), o-BUDTMS (blue), and o-DTMS (black) oligomers. Solid curves and dashed lines are VFTH and Arrhenius fits to the experimental points, respectively. Filled point is $T_g$ value calculated from MT-DSC measurements (period = 60 s, τ = ~ 10 s).

Whereas o-OTS exhibits a heat capacity step around 245 K, no step-like change could be assuredly distinguished on TM-DSC curves of o-DTMS and o-BUDTMS. As for monomers, the reason of this discrepancy might be the higher crystallinity ratios of o-DTMS and o-BUDTMS systems due to their shorter alkyl chains. Nevertheless, the glass transition temperature of o-OTS estimated by TM-DSC is in close agreement with dielectric spectroscopy measurements.

**Molecular mobility of self-assembled monolayers**

Self-assembled monolayers (SAMs) are highly ordered monomolecular height layers formed by the spontaneous adsorption of molecules on a solid surface. In such nanolayers, molecule headgroups are firmly attached to the substrate, thus forming the anchoring zone, while intermolecular interactions held molecular tails in an upward position near the substrate normal. Particularly, SAMs formed from monopolar monomers (OTS and DTMS) possess a polar anchoring zone whereas monolayers of bipolar molecules (BUDTMS) have an additional polar canopy.

The quality of the self-assembled monolayers was investigated by atomic force microscopy (AFM), ellipsometry and water contact angle (WCA) measurements. The AFM images (Fig. 5) reveal relatively smooth surfaces for OTS and DTMS SAMs, although a few pinholes are apparent. The depths of these holes (measured between blue triangles in Fig. 5) are remarkably close to the calculated and ellipsometry-determined monolayer thicknesses (Tab. 1). Thus, such pinholes were attributed to locations where a few molecules are not grafted to the silicon substrate. Silicon substrates being totally flat and finely polished, the dark straight lines observed on OTS and DTMS images likely originate from the wiping of sample surfaces applied to remove molecular aggregates formed during sample preparation.[33] However, number of these aggregates is observed on the BUDTMS monolayer (Fig. 5c) even after sustained wiping. Strong affinity of these molecular aggregates to the underlying SAM likely originates from electrostatic interactions between dipolar groups contained in aggregates and bromine endgroups forming the monolayer surface. Ellipsometric and water contact angle measurements were performed to further confirm the thicknesses and homogeneity of monolayers. Results are presented in Table 1. The monolayer thicknesses determined by both AFM and ellipsometry were found to be in close agreement with thicknesses estimated by theoretical calculations. Moreover, WCAs measured on monolayers are also in agreement with results reported earlier.[33,75,76] Thus, ellipsometry, AFM and WCAs experimental results all indicate homogeneous monolayers with good quality, even though the polar bromine-termination of the BUDTMS precursors might slightly disturb the assembly process and led to the formation of small aggregates increasing the RMS value of the BUDTMS SAMs.

The mobility of molecules within SAMs formed from OTS, DTMS, and BUDTMS were then investigated by dielectric spectroscopy. Although it could be expected that all building blocks undergo one relaxation process of their alkylsilane headgroups and that BUDTMS manifests a second relaxation signal corresponding to the motions of bromoalkyl endgroups, dielectric experiments revealed that only BUDTMS monolayers have an observable relaxation process (Fig. 6a). No significant signals due to a relaxation process could be clearly observed in spectra measured for DTMS and OTS monolayers (see supporting information, SI-4 and -5). These results indicate that the relaxation process of alkylsilane headgroups cannot be observed under our experimental conditions. Three possible causes may explain the absence of alkylsilane molecular relaxations in BDS spectra: (1) the relaxation signal is too weak to be distinguished from experimental noise, (2) the molecular relaxations occur at temperatures out of the temperature range that can be reached by the electrodes (i.e. 110 to 440 K), (3) the mobile segments of dipolar headgroups are highly constrained by the structure of the monolayer and, therefore, unable to fluctuate. This latter case is the most plausible one as both the rigid covalent binding of the silane headgroups to the substrate and tight packing of molecules within the SAMs likely prevent these dipolar headgroups to reorient.

On the contrary, the flexibility of the alkyl chains may allow a certain degree of flexibility to the molecular segments distant from the silane anchoring points, and more especially to the monolayer canopy. Thus, the dipolar groups at the termination of building blocks, such as bromoalkyl endgroups in BUDTMS, are able to fluctuate and their relaxation signal can be clearly observed. This relaxation signal could also originate from the motions of molecules within small aggregates on top of the SAMs as observed on AFM images. However, building blocks forming such aggregates are fairly free to move as they are loosely-packed and not covalently attached to the substrate, therefore, a relaxation signal corresponding to the motions of alkylsilane headgroups, similar to the one observed in oligomers, would be clearly observable on BDS spectra. Consequently, it can be assumed that the contribution of these aggregates to the observed molecular relaxation is trivial and that the observed relaxation process of BUDTMS monolayers can be solely attributed to the motions of bromoalkyl endgroups forming the monolayer canopy.

The spectra have been analyzed by fitting the HN-equation (Equation (4)) to the data (see Fig. 6b) including a conductivity contribution to the dielectric loss according to $\sigma/[\varepsilon_0(2\pi f)^n]$ where $\sigma$ is related to the conductivity and $n$ an exponent. As depicted in Fig. 7a, the temperature dependence of the relaxation time is curved when plotted versus 1/T which might indicate that molecule endgroups move cooperatively. This cooperativity possibly arises from the inherent close-packed structure of the monolayers that promotes interactions between adjacent building blocks, and more specifically, from the electrostatic forces between bromoalkyl endgroups. Hence, tight molecular packing and significant intermolecular interactions also raise the energy barrier required for the motions of endgroups which leads to a higher dielectric glass transition temperature ($T_\alpha$) of 266 K.

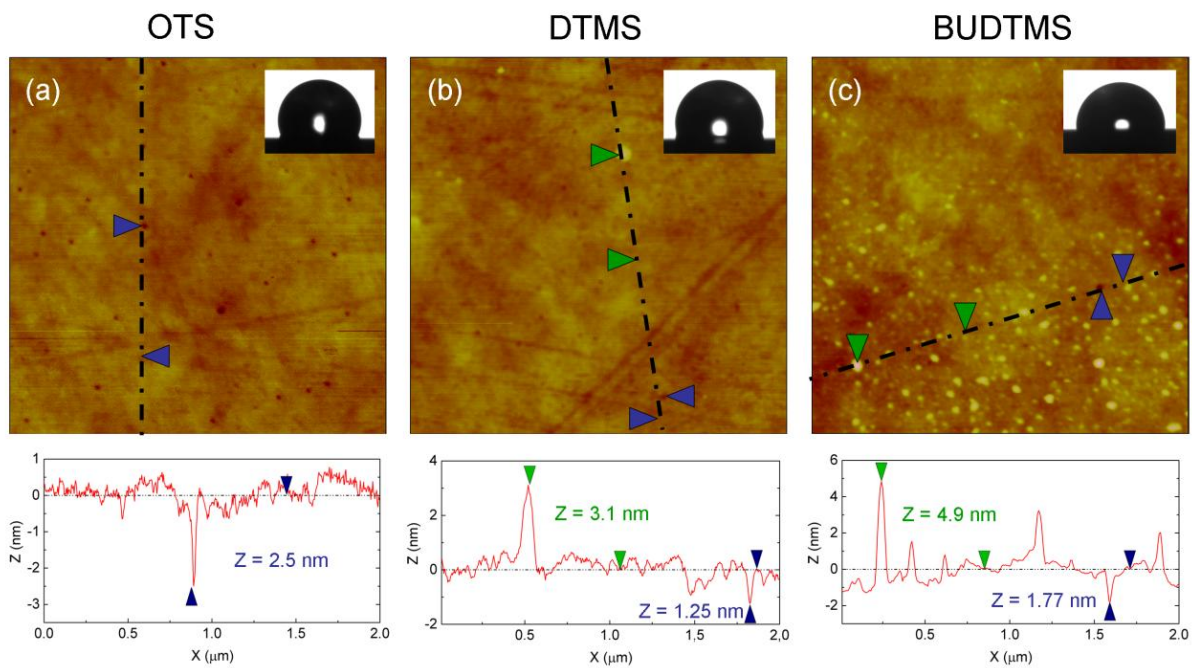

**Fig 5.** AFM images of (a) OTS, (b) DTMS, and (c) BUDTMS self-assembled monolayers. Scan area and height are 2 x 2 μm² and 0-10 nm, respectively. Height profiles are represented along the dashed-doted black lines on AFM images. Blue and green triangles correspond to the locations where the heights were measured. (Inset) images of a water droplet on each monolayer.

The cooperative character of this relaxation was further investigated employing the concept proposed by Eyring and revised by Starkweather.[77,78] This approach uses the activation entropy ($\Delta S^*$) calculated from Eq. 8 as a measure for the cooperativity of molecular motions.

$$\Delta S^* = \frac{\Delta H^* - \Delta H_0}{T} \quad (8)$$

where $\Delta H^*$ is the activation enthalpy. Under isobaric conditions $\Delta H^*$ can be replaced by a temperature dependent activation energy which can be estimated from the relaxation map (see Fig. 7a). $\Delta H_0$ corresponds to the theoretical activation energy when $\Delta S^* = 0$ (Eq. 9).

$$\Delta H_0 = RT \left[1 + ln\left(\frac{k_b T}{2\pi h f}\right)\right] \quad (9)$$

where $k_b$ is the Boltzmann constant and $h$ the Planck constant. Thus, molecular relaxations whose activation entropy values $\Delta S^*$ are greater than zero, or $\Delta H^*$ values are higher than $\Delta H_0$, can be considered as cooperative. According to these considerations, Fig. 7b reveals that the molecular fluctuations of bromoalkyl dipoles in BUDTMS monolayers are cooperative. Indeed, the activation entropy ($\Delta S^*$) associated to this relaxation reaches a maximum value of 0.25 kJ.mol⁻¹.K⁻¹ at low temperatures. Even though the reorientation of bromoalkyl dipoles would be apparently local, the activation entropy of this relaxation is remarkably greater than that expected for secondary local $\beta$ processes,[79,80] and is rather comparable to larger scale molecular motions associated to $\alpha$ processes.[81,82] Such a high cooperativity suggests that the relaxation of bromoalkyl endgroups structurally corresponds to the collective motions of large parcels of the monolayer canopy. Fig. 7b also reveals that the cooperativity of motions progressively decreases when temperature is increased indicating that local molecular motions are facilitated by the increase in the free volume fraction within the material.[83] It is worth noting that theoretical activation enthalpy values shown in Fig. 7b are calculated at different frequencies with increasing values. For this reason, these values tend to decrease with temperature in Fig. 7b while they are expected to increase (see supporting information, SI-7). Overall, the high transition temperature and cooperative nature of this relaxation highlight the significant amount of energy required for this relaxation to occur.

**Tab 1.** Thickness, water contact angle and RMS values of OTS, DTMS and BUDTMS monolayers.

|  | Monolayer thickness (nm) | | | Water contact angle (°) | | RMS (nm) |
|---|---|---|---|---|---|---|
|  | Calculated[a] | Ellipsometry[b] | AFM | Measured | References |  |
| OTS | 2.43 | 2.62 | 2.5 | 110 ± 1.6 | 109 (Ref.[75]) | 0.21 |
| DTMS | 1.68 | 1.62 | 1.25 | 104 ± 1.4 | 105 (Ref.[33]) | 0.22 |
| BUDTMS | 1.66 | 1.52 | 1.77 | 85 ± 1.2 | 85 (Ref.[76]) | 0.49 |

[a] thicknesses are calculated considering the length of an all-trans molecule and assuming a tilt angle of 25° from the surface normal. [b] an error of 0.2 nm is assumed.

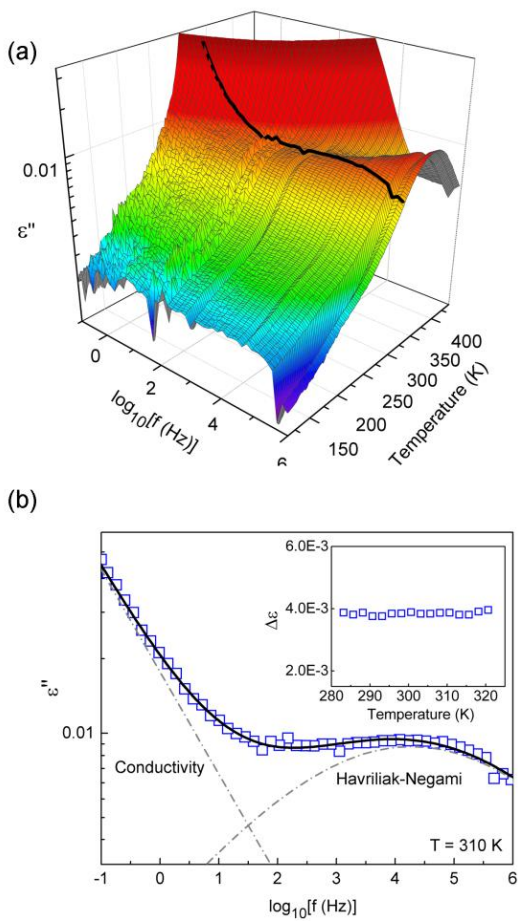

**Fig 6.** (a) Dielectric loss (ε") of a BUDTMS monolayer as a function of frequency and temperature. (b) Dielectric loss (ε") data corresponding to the black curve on spectra (a) as a function of frequency at a fixed temperature (T = 310 K). Dashed-dotted grey curves are the contributions of the relaxation process described by a Havriliak-Negami function and that of the conductivity. The black solid line corresponds to the resulting fit to data including both contributions. Inset shows the very weak variation of the dielectric strength Δε with temperature.

Further information concerning the polarizability and cooperativity of this system were revealed by using the autocorrelation function. Time-dependent correlation functions are useful in dielectric spectroscopy studies to probe the total electrical polarization of the system. Information about molecular reorientation, such as the distribution of relaxation times, can then be plotted as a function of time. By using the estimated HN parameters the dielectric response can be calculated in the time domain,[84,85] as shown in Fig. 8. In the case of the bromoalkyl relaxation in BUDTMS monolayers, the distribution of relaxation times is remarkably broad covering the range from $10^{-11}$ to $10^5$ s. Time-correlation curves shown in Fig. 8 were fitted by the empirical Kohlrausch-Williams-Watts (KWW) stretched exponential function (Eq. 12) expressed as follows:

$$\phi(t) = \exp\left[-\left(\frac{t}{\tau}\right)^{\beta_{KWW}}\right], \quad 0 < \beta_{KWW} < 1 \quad (10)$$

where $\phi(t)$ is the time-relaxation function, $t$ the time, $\tau$ the characteristic relaxation time, and $\beta_{KWW}$ a shape parameter that accounts for the stretching character of the exponential function. In this expression, $\beta_{KWW}$ represents the dispersive character of $\phi(t)$ so that high values suggest that a unique relaxation time $\tau$ is enough to describe the correlation function, while low $\beta_{KWW}$ values indicate a large dispersion between the slowest and fastest relaxation times.[86] In the case of the α process in BUDTMS SAMs, the value of $\beta_{KWW}$ was found to slightly increase with temperature and an average value of $\beta_{KWW} = 0.13$ was estimated (Fig. 8, inset). This value is remarkably low compared to values commonly encountered for α processes of polymeric glass-formers[87,88] and suggests a strongly heterogeneous relaxation where some fast molecular units relax well before other slow units. It also supports the proposition stating that local structural steric constraints and numerous intermolecular interactions significantly hinder the motions of bromoalkyl dipoles.[89] Consequently, the independent motion of individual bromoalkyl endgroups is unlikely. Instead, large parcels of the monolayer canopy move cooperatively. As depicted in Fig. 8, these parcels progressively shrink due to a decrease of the cooperativity with temperature. Overall, the presence of a dipolar endgroup in BUDTMS molecule has allowed us to easily track molecular motions within the monolayer canopy by dielectric spectroscopy. However, the bulkiness of these groups may have also induced some structural defects or a more loosely-packed canopy thus giving a greater flexibility to this portion of SAMs. Consequently, molecular dynamics observed within BUDTMS SAM may not accurately render the mobility of more conventional monolayers composed of highly ordered monopolar molecules. For this reason, further investigations were conducted on conductivity contributions. Conductivity signals arise from the motions of charge carriers in a material,[90] and, in the case of sinusoidal electric fields, can be estimated through the measure of the complex dielectric function $\varepsilon^*$ as follows:[91]

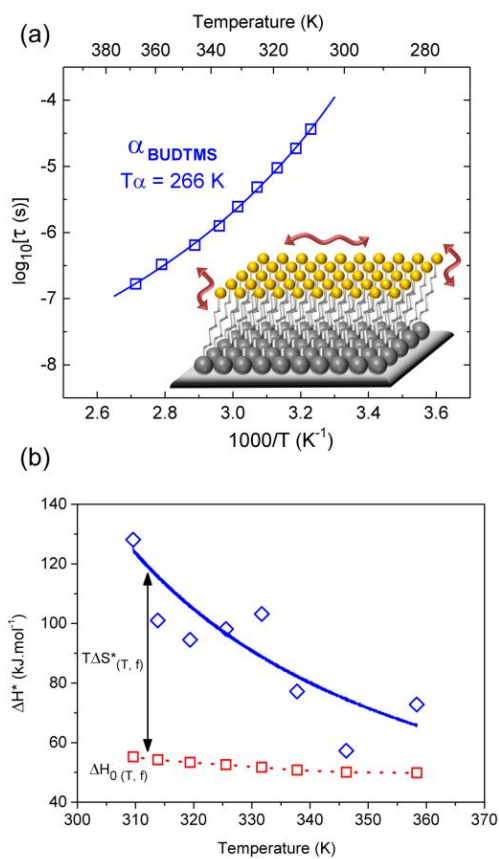

**Fig 7.** (a) Relaxation map for the BUDTMS self-assembled monolayer. The solid curve corresponds to a VFTH fit to the data. The inset illustrates the motions of the bromoalkyl dipolar endgroups forming the canopy of a BUDTMS self-assembled monolayer. (b) Activation enthalpy ($\Delta H^*$) (blue diamonds) of the BUDTMS monolayer relaxation as a function of temperature. The solid line corresponds to the calculation of $\Delta H^*$ from the VFTH fit. The values of $\Delta H_0$ are also represented as a function of temperature (red squares). The dotted red line is a guide to the eyes.

$$\sigma^*(f) = i2\pi f \varepsilon_0 \varepsilon^* \qquad (11)$$

where $\varepsilon_0$ is the permittivity of the free space. As shown in Fig. 9a, at low temperatures, the logarithm value of $\sigma'$, the real part of the conductivity, nearly follows a linear increase with frequency. As temperature is increased, a plateau at low frequencies (d.c. conductivity) progressively appears whereas the slope at high frequencies (a.c. conductivity) remains unchanged. According to J.C. Dyre, this evolution of the conductivity can be universally predicted and would be uniquely dependent on the frequency.[92]

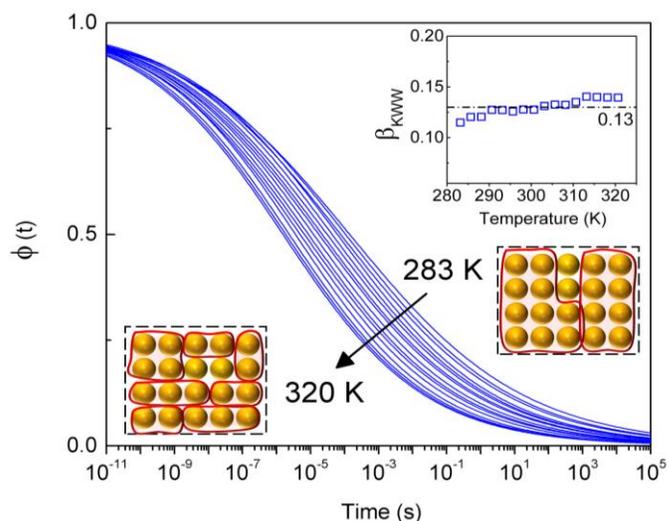

**Fig 8.** Time dependence of the correlation function due to the dielectric loss of BUDTMS SAMs. The inset plot shows the variation of the $\beta_{KWW}$ parameter as a function of temperature. Inset drawings represent top views of BUDTMS SAMs with cooperative regions depicted by red areas.

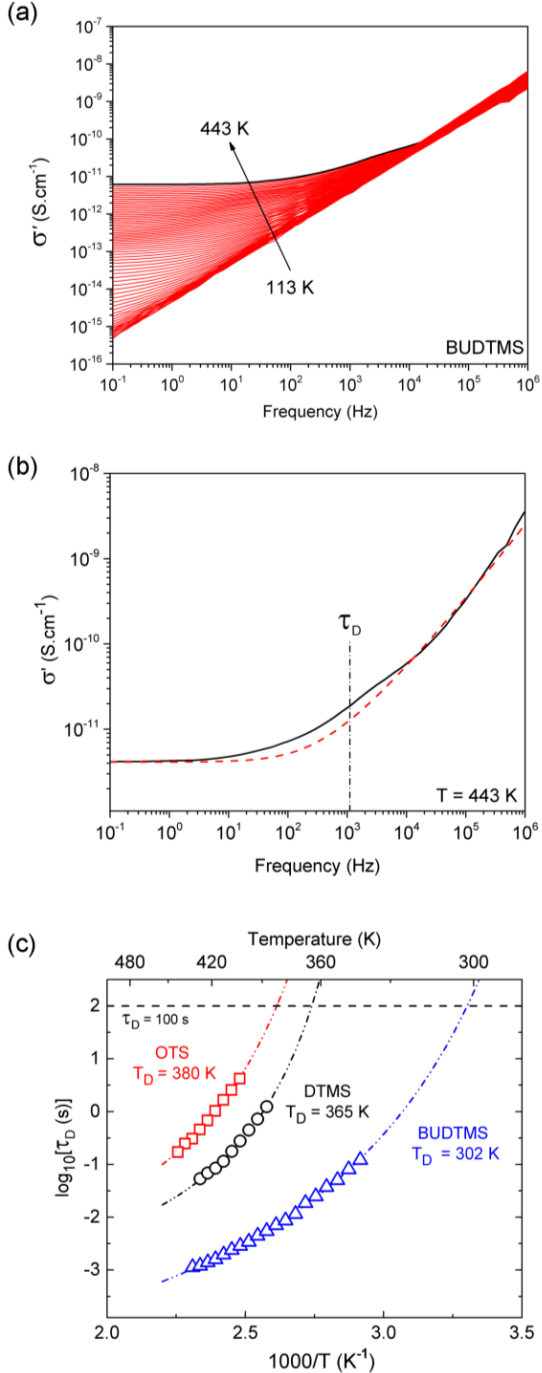

**Fig 9.** (a) Real part of the conductivity (σ') of the BUDTMS monolayer as a function of frequency and temperature. The black curve corresponds to the conductivity signal at 443 K. (b) Experimental (solid black) and fit to data from Dyre's model (dashed red) curves. τ$_D$ corresponds to the transition from DC to AC conductivity. (c) Relaxation maps of BUDTMS (blue triangles), DTMS (black circles) and OTS (red squares) constructed from values of τ$_D$ at different temperatures. Dash-dotted curves correspond to VFTH fits. $T_D$ are temperatures at which fits reach $\tau_D = 100s$.

Thus, a transition point, $\tau_D$, can be estimated at the transition from d.c. to a.c. conductivity from the following equation:

$$\sigma'(f) = \sigma_0 \frac{2\pi f \tau_D \arctan(2\pi f \tau_D)}{\left\{ln[1+(2\pi f \tau_D)^2]^{\frac{1}{2}}\right\}^2 + [\arctan(2\pi f \tau_D)]^2} \quad (12)$$

where $\sigma_0$ is the d.c. conductivity. Hence, $\sigma'$ can be plotted as a function of frequency and fitted to data to determine values of $\tau_D$ at different temperatures (Fig. 9b). Thus, as previously depicted, relaxation maps can be made. Fig. 9c shows the variation of $\tau_D$ as a function of temperature for BUDTMS, DTMS and OTS monolayers. Although BUDTMS and DTMS molecules are very similar, Fig. 9c highlights the higher transition temperature (measured at $\tau_D = 100\ s$) of the DTMS monolayer ($T_D(DTMS) = 365\ K$) compared to the BUDTMS SAM ($T_D(BUDTMS) = 302\ K$). This higher temperature can be attributed to a higher ordering level of molecules in the DTMS monolayer. Indeed, the bulky endgroups formed by bromine atoms at the end of BUDTMS molecules are likely to slightly disrupt the organization of molecules and particularly decrease the ordering of molecular units within the monolayer canopy. Therefore, the whole monolayer is more flexible and requires less energy to move, thus leading to a lower transition temperature. Interestingly, this result indicates that even though numerous physical interactions exist between adjacent endgroups, which could have led to a higher cohesion of molecules in the monolayer, their poor ordering significantly increases the ability of SAM building blocks to move. Accordingly, the OTS monolayer exhibits a higher transition temperature ($T_D(OTS) = 380\ K$). Intrinsically, the long methyl-terminated chain of OTS is expected to increase the monolayer level of ordering,[93,94] thus leading to a higher transition temperature. In consequence, Fig. 9c also evidences the decrease of the relaxation time ($\tau_D$) at a fixed temperature when the molecular ordering lowers.

More generally, two parameters should be carefully examined when choosing building blocks for self-assembled monolayers: molecular steric hindrance and potential interactions between molecules. According to the application, molecules containing bulky groups may be preferred to increase molecular mobility, whereas building blocks designed to increase intermolecular interactions would lead to a lower mobility of molecules.

## Conclusion

The effect of structural constriction on molecular mobility was investigated by broadband dielectric spectroscopy within three types of alkylsilane-containing self-assembled nanostructures. Molecular fluctuations were analyzed as a function of frequency, temperature, nanostructures and building block composition. Although all precursor monomers exhibit cooperative motions of their alkylsilane headgroups due to strong intermolecular interactions, the bromine-termination of bipolar monomers manifests a distinct relaxation process corresponding to the local reorientations of bromoalkyl dipolar units.

A totally different molecular mobility was observed after formation of oligomers. Polysiloxane oligomers resulting from the condensation of monomers were found to have a phase-separated nanostructure where alkyl side chains form nanodomains independent from the siloxane main chain. While main chains of these oligomers undergo non-cooperative motions, independent nanodomains of side chains were found to relax cooperatively due to numerous intermolecular interactions.

Remarkably, the highly ordered and constrained nanostructure of self-assembled monolayers greatly modified the relaxation processes of both alkylsilane and bromoalkyl dipoles. The close-packed structure and rigid covalent attachment of alkylsilane headgroups to the substrate prevented the experimental observation of their fluctuations. Interestingly, this limitation allowed observing and accurately analyzing motions of bromoalkyl endgroups forming the monolayer canopy independently from other overlapping dielectric signals. This relaxation process was revealed as being strongly cooperative due to both the tight molecular packing of SAMs that creates local structural steric constraints, and the presence of abundant intermolecular interactions between adjacent polar endgroups. These structural constraints thus led to a heterogeneous relaxation with a large dispersion of relaxation times.

Finally, these findings not only support several previously published results[31,32] but also broaden the understanding of processes governing motions of molecules within self-assembled monolayers. They could be greatly appealing to induce dynamic properties to functional coating materials. Hypothetically, they would also encourage the development of novel stimuli-responsive nanotechnologies or actuators by taking advantage of these dynamic and reversible molecular motions.

## Acknowledgements

Authors are grateful to the partial financial support from the French Department for Higher Education and Research, and the EU-US Transatlantic Degree Program in Engineering. They also acknowledge the Upper Normandy region for funding the broadband dielectric spectrometer. LT appreciates the financial support from National Science Foundation (CMMI-1068952 & IIA-1338988) and Nebraska Research Initiative. Morgane Sanselme from the SMS laboratory (University of Rouen) and Jacques Loubens from TA Instruments are greatly acknowledged for their generous help.

## References

1   A. S. Eggeman, S. Illig, A. Troisi, H. Sirringhaus and P. A. Midgley, *Nat. Mater.*, 2013, **12**, 1045–1049.
2   Z. Chen, G. Wang, Z. Xu, H. Li, A. Dhôtel, X. C. Zeng, B. Chen, J.-M. Saiter and L. Tan, *Adv. Mater.*, 2013, **25**, 6106–6111.
3   D. Y. Zhang and G. Seelig, *Nat. Chem.*, 2011, **3**, 103–113.
4   A. Dhotel, L. Delbreilh, B. Youssef, J. Jiang, G. Coquerel, J.-M. Saiter and L. Tan, *J. Therm. Anal. Calorim.*, 2013, **112**, 301–305.
5   J. Bath and A. J. Turberfield, *Nat. Nanotechnol.*, 2007, **2**, 275–284.
6   M. A. C. Stuart, W. T. S. Huck, J. Genzer, M. Müller, C. Ober, M. Stamm, G. B. Sukhorukov, I. Szleifer, V. V. Tsukruk, M. Urban, F. Winnik, S. Zauscher, I. Luzinov and S. Minko, *Nat. Mater.*, 2010, **9**, 101–113.
7   A. Dhotel, Z. Chen, L. Delbreilh, B. Youssef, J.-M. Saiter and L. Tan, *Int. J. Mol. Sci.*, 2013, **14**, 2303–2333.
8   R. Arcos-Ramos, B. Rodríguez-Molina, M. Romero, J. M. Méndez-Stivalet, M. E. Ochoa, P. I. Ramírez-Montes, R. Santillan, M. A. Garcia-Garibay and N. Farfán, *J. Org. Chem.*, 2012, **77**, 6887–6894.
9   G. T. Carroll, G. London, T. F. Landaluce, P. Rudolf and B. L. Feringa, *ACS Nano*, 2011, **5**, 622–630.
10  H. M. D. Bandara and S. C. Burdette, *Chem. Soc. Rev.*, 2012, **41**, 1809–1825.
11  Z. Pei, Y. Yang, Q. Chen, E. M. Terentjev, Y. Wei and Y. Ji, *Nat. Mater.*, 2014, **13**, 36–41.
12  K. Arabeche, L. Delbreilh, R. Adhikari, G. H. Michler, A. Hiltner, E. Baer and J.-M. Saiter, *Polymer*, 2012, **53**, 1355–1361.
13  K. Arabeche, L. Delbreilh, J.-M. Saiter, G. H. Michler, R. Adhikari and E. Baer, *Polymer*, 2014, **55**, 1546–1551.
14  J. M. Mativetsky, G. Pace, M. Elbing, M. A. Rampi, M. Mayor and P. Samorì, *J. Am. Chem. Soc.*, 2008, **130**, 9192–9193.
15  N. Crivillers, E. Orgiu, F. Reinders, M. Mayor and P. Samorì, *Adv. Mater.*, 2011, **23**, 1447–1452.
16  R. Madueno, M. T. Räisänen, C. Silien and M. Buck, *Nature*, 2008, **454**, 618–621.
17  L. Tauk, A. P. Schröder, G. Decher and N. Giuseppone, *Nat. Chem.*, 2009, **1**, 649–656.
18  J. Jiang, O. V. Lima, Y. Pei, Z. Jiang, Z. Chen, C. Yu, J. Wang, X. C. Zeng, E. Forsythe and L. Tan, *ACS Nano*, 2010, **4**, 3773–3780.
19  M. Halik and A. Hirsch, *Adv. Mater. Deerfield Beach Fla*, 2011, **23**, 2689–2695.
20  T. Minari, C. Liu, M. Kano and K. Tsukagoshi, *Adv. Mater.*, 2012, **24**, 299–306.

21 M. F. Calhoun, J. Sanchez, D. Olaya, M. E. Gershenson and V. Podzorov, *Nat. Mater.*, 2008, **7**, 84–89.
22 J. Shao, E. A. Josephs, C. Lee, A. Lopez and T. Ye, *ACS Nano*, 2013, **7**, 5421–5429.
23 J. A. Bardecker, A. Afzali, G. S. Tulevski, T. Graham, J. B. Hannon and A. K.-Y. Jen, *Chem. Mater.*, 2012, **24**, 2017–2021.
24 Y. Li, C.-Y. Xu, P. Hu and L. Zhen, *ACS Nano*, 2013, **7**, 7795–7804.
25 S. Ido, H. Kimiya, K. Kobayashi, H. Kominami, K. Matsushige and H. Yamada, *Nat. Mater.*, 2014, **13**, 264–270.
26 S. T. Marshall, M. O'Brien, B. Oetter, A. Corpuz, R. M. Richards, D. K. Schwartz and J. W. Medlin, *Nat. Mater.*, 2010, **9**, 853–858.
27 C. J. Barile, E. C. M. Tse, Y. Li, T. B. Sobyra, S. C. Zimmerman, A. Hosseini and A. A. Gewirth, *Nat. Mater.*, 2014, **13**, 619–623.
28 G. T. Carroll, G. London, T. F. Landaluce, P. Rudolf and B. L. Feringa, *ACS Nano*, 2011, **5**, 622–630.
29 M. Elbing, A. Błaszczyk, C. von Hänisch, M. Mayor, V. Ferri, C. Grave, M. A. Rampi, G. Pace, P. Samorì, A. Shaporenko and M. Zharnikov, *Adv. Funct. Mater.*, 2008, **18**, 2972–2983.
30 G. B. Demirel, N. Dilsiz, M. Çakmak and T. Çaykara, *J. Mater. Chem.*, 2011, **21**, 3189–3196.
31 Q. Zhang, Q. Zhang and L. A. Archer, *J. Phys. Chem. B*, 2006, **110**, 4924–4928.
32 M. C. Scott, D. R. Stevens, J. R. Bochinski and L. I. Clarke, *ACS Nano*, 2008, **2**, 2392–2400.
33 Y. Ito, A. A. Virkar, S. Mannsfeld, J. H. Oh, M. Toney, J. Locklin and Z. Bao, *J. Am. Chem. Soc.*, 2009, **131**, 9396–9404.
34 G. Schaumburg, *Dielectr. Newsl.*, 2006, **22**, 5–7.
35 C. M. Herzinger, B. Johs, W. A. McGahan, J. A. Woollam and W. Paulson, *J. Appl. Phys.*, 1998, **83**, 3323–3336.
36 F. Kremer and A. Schönhals, *Broadband Dielectric Spectroscopy*, Springer Berlin Heidelberg, Berlin Heidelberg, 2003.
37 F. Kremer and A. Schönhals, in *Broadband Dielectric Spectroscopy*, eds. P. D. F. Kremer and P.-D. D. A. Schönhals, Springer Berlin Heidelberg, 2003, pp. 99–129.
38 U. Schneider, P. Lunkenheimer, A. Pimenov, R. Brand and A. Loidl, *Ferroelectrics*, 2001, **249**, 89–98.
39 A. Saiter, L. Delbreilh, H. Couderc, K. Arabeche, A. Schönhals and J.-M. Saiter, *Phys. Rev. E*, 2010, **81**, 041805.
40 A. Schönhals, H. Goering, C. Schick, B. Frick and R. Zorn, *Eur. Phys. J. E*, 2003, **12**, 173–178.
41 V. M. Boucher, D. Cangialosi, H. Yin, A. Schönhals, A. Alegría and J. Colmenero, *Soft Matter*, 2012, **8**, 5119–5122.
42 R. Crétois, L. Delbreilh, E. Dargent, N. Follain, L. Lebrun and J. M. Saiter, *Eur. Polym. J.*, 2013, **49**, 3434–3444.
43 N. Delpouve, L. Delbreilh, G. Stoclet, A. Saiter and E. Dargent, *Macromolecules*, 2014.
44 A. Dhotel, B. Rijal, L. Delbreilh, E. Dargent and A. Saiter, *J. Therm. Anal. Calorim.*, 2014, Submitted.
45 H. K. Nguyen, D. Prevosto, M. Labardi, S. Capaccioli, M. Lucchesi and P. Rolla, *Macromolecules*, 2011, **44**, 6588–6593.
46 B. Vanroy, M. Wübbenhorst and S. Napolitano, *ACS Macro Lett.*, 2013, **2**, 168–172.
47 Q. Tang, W. Hu and S. Napolitano, *Phys. Rev. Lett.*, 2014, **112**, 148306.
48 K. Watanabe, T. Kawasaki and H. Tanaka, *Nat. Mater.*, 2011, **10**, 512–520.
49 H. Oh and P. F. Green, *Nat. Mater.*, 2009, **8**, 139–143.
50 H. Yin, S. Napolitano and A. Schönhals, *Macromolecules*, 2012, **45**, 1652–1662.
51 D. Cangialosi, in *Dynamics in Geometrical Confinement*, ed. F. Kremer, Springer International Publishing, 2014, pp. 339–361.
52 S. Haldar, K. C. Dey, D. Sinha, P. K. Mandal, W. Haase and P. Kula, *Liq. Cryst.*, 2012, **39**, 1196–1203.
53 Ł. Kolek, M. Massalska-Arodź, M. Paluch, K. Adrjanowicz, T. Rozwadowski and D. Majda, *Liq. Cryst.*, 2013, **40**, 1082–1088.
54 *Soft Matter*.
55 S. Havriliak and S. Negami, *Polymer*, 1967, **8**, 161–210.
56 P. J. Purohit, J. E. Huacuja-Sánchez, D.-Y. Wang, F. Emmerling, A. Thünemann, G. Heinrich and A. Schönhals, *Macromolecules*, 2011, **44**, 4342–4354.
57 L. Hartmann, F. Kremer, P. Pouret and L. Léger, *J. Chem. Phys.*, 2003, **118**, 6052.
58 A. R. Bras, J. P. Noronha, M. M. Cardoso, A. Schönhals, F. Affouard, M. Dionísio and N. T. Correia, *J. Phys. Chem. B*, 2008, **112**, 11087–11099.
59 H. Vogel, *Phys Z*, 1921, **22**, 645–646.
60 G. S. Fulcher, *J. Am. Ceram. Soc.*, 1925, **8**, 339–355.
61 G. Tammann and W. Hesse, *Z. Für Anorg. Allg. Chem.*, 1926, **156**, 245–257.
62 C. M. Roland, *Soft Matter*, 2008, **4**, 2316–2322.
63 J. F. Mano and S. Lanceros-Méndez, *J. Appl. Phys.*, 2001, **89**, 1844–1849.
64 A. Dhotel, H. Li, L. Fernandez-Ballester, L. Delbreilh, B. Youssef, X. C. Zeng and L. Tan, *J. Phys. Chem. C*, 2011, **115**, 10351–10356.
65 F. D. Osterholtz and E. R. Pohl, *J. Adhes. Sci. Technol.*, 1992, **6**, 127–149.
66 A. Dhotel, Z. Xu, L. Delbreilh, B. Youssef, J. M. Saiter and L. Tan, *MATEC Web Conf.*, 2013, **3**, 4.
67 M. Beiner, K. Schröter, E. Hempel, S. Reissig and E. Donth, *Macromolecules*, 1999, **32**, 6278–6282.
68 M. Beiner and H. Huth, *Nat. Mater.*, 2003, **2**, 595–599.
69 S. Pankaj and M. Beiner, *Soft Matter*, 2010, **6**, 3506–3516.
70 S. Pankaj and M. Beiner, *J. Phys. Chem. B*, 2010, **114**, 15459–15465.
71 X. Shi, D. Graiver and R. Narayan, *Silicon*, 2012, **4**, 109–119.
72 M. Brand, A. Frings, P. Jenker, R. Lehnert, H. J. Metternich, J. Monkiewicz and J. Schram, *Z Naturforsch 54b*, 1999, **155**.
73 R. Lam and P. H. Geil, *Polym. Bull.*, 1978, **1**, 127–131.

74 U. Gaur and B. Wunderlich, *Macromolecules*, 1980, **13**, 445–446.
75 M.-H. Jung and H.-S. Choi, *Korean J. Chem. Eng.*, 2009, **26**, 1778–1784.
76 C. Haensch, S. Hoeppener and U. S. Schubert, *Chem. Soc. Rev.*, 2010, **39**, 2323–2334.
77 H. W. Starkweather, *Macromolecules*, 1981, **14**, 1277–1281.
78 H. W. Starkweather, *Macromolecules*, 1988, **21**, 1798–1802.
79 F. Meersman, B. Geukens, M. Wübbenhorst, J. Leys, S. Napolitano, Y. Filinchuk, G. Van Assche, B. Van Mele and E. Nies, *J. Phys. Chem. B*, 2010, **114**, 13944–13949.
80 S. Sharifi, D. Prevosto, S. Capaccioli, M. Lucchesi and M. Paluch, *J. Non-Cryst. Solids*, 2007, **353**, 4313–4317.
81 J. F. Mano, *Macromolecules*, 2001, **34**, 8825–8828.
82 A.-C. Genix and F. Lauprêtre, *Macromolecules*, 2005, **38**, 2786–2794.
83 S. Matsuoka, G. H. Fredrickson and G. E. Johnson, in *Molecular Dynamics and Relaxation Phenomena in Glasses*, eds. T. Dorfmüller and G. Williams, Springer Berlin Heidelberg, 1987, pp. 188–202.
84 F. Alvarez, A. Alegra and J. Colmenero, *Phys. Rev. B*, 1991, **44**, 7306–7312.
85 G. Williams, *Chem. Rev.*, 1972, **72**, 55–69.
86 L. Berthier, G. Biroli, J.-P. Bouchaud, L. Cipelletti and W. van Saarloos, *Dynamical Heterogeneities in Glasses, Colloids, and Granular Media*, Oxford University Press, 2011.
87 K. L. Ngai, *J. Phys. Condens. Matter*, 2003, **15**, S1107.
88 Y. Guo, C. Zhang, C. Lai, R. D. Priestley, M. D'Acunzi and G. Fytas, *ACS Nano*, 2011, **5**, 5365–5373.
89 K. L. Ngai, *Relaxation and Diffusion in Complex Systems*, Springer, New York, 2011.
90 P. Lunkenheimer and A. Loidl, in *Broadband Dielectric Spectroscopy*, eds. P. D. F. Kremer and P.-D. D. A. Schönhals, Springer Berlin Heidelberg, 2003, pp. 131–169.
91 F. Kremer and S. A. Różański, in *Broadband Dielectric Spectroscopy*, eds. P. D. F. Kremer and P.-D. D. A. Schönhals, Springer Berlin Heidelberg, 2003, pp. 475–494.
92 J. C. Dyre, *Phys. Lett. A*, 1985, **108**, 457–461.
93 A. Ulman, *Chem. Rev.*, 1996, **96**, 1533–1554.
94 F. Schreiber, *Prog. Surf. Sci.*, 2000, **65**, 151–257.